\documentclass{article}
\usepackage{spconf,amsmath,graphicx}
\usepackage{xspace}
\usepackage[smallerops]{newtxmath}
\usepackage{csquotes}
\usepackage{todonotes}
\usepackage{nicefrac}
\usepackage{url}
\usepackage{listings}
\usepackage{booktabs}
\usepackage{tablefootnote}
\usepackage{bbm}
\usepackage{multicol}
\usepackage{adjustbox}
\usepackage{xcolor}
\usepackage{verbatim}

\title{AN ASYNCHRONOUS WFST-BASED DECODER FOR AUTOMATIC SPEECH RECOGNITION}
\name{Hang Lv$^{1,2}$, Zhehuai Chen$^{2,5}$, Hainan Xu$^2$,  Daniel Povey$^4$, Lei Xie$^1$, Sanjeev Khudanpur$^{2,3}$}
\address{
\textsuperscript{1}~Audio, Speech and Language Processing Lab (ASLP@NPU), \\School of Computer Science, Northwestern Polytechnical University, Xi'an, China\\
\textsuperscript{2}~Center of Language and Speech Processing,
\textsuperscript{3}~Human Language Technology Center of Excellence, \\Johns Hopkins University, Baltimore, MD, USA \\
\textsuperscript{4}~Xiaomi Corporation, Beijing, China \\
\textsuperscript{5}~SpeechLab, Department of Computer Science and Engineering, Shanghai Jiao Tong University\\
\small{\texttt{\string{hanglv,lxie\string}@nwpu-aslp.org},
\texttt{chenzhehuai@sjtu.edu.cn},
\texttt{\string{hxu31,khudanpur\string}@jhu.edu}, \texttt{dpovey@xiaomi.com}}
\thanks{Lei Xie is the corresponding author. The codes associated with this work can be found from \protect\url{https://github.com/LvHang/kaldi/tree/async-a-star-decoder}}
}

\begin{document}
\ninept

\maketitle

\begin{abstract}
We introduce \emph{asynchronous dynamic decoder}, which adopts an efficient A* algorithm to incorporate big language models in the one-pass decoding for large vocabulary continuous speech recognition.
Unlike standard one-pass decoding with on-the-fly composition decoder which might induce a significant
computation overhead, the asynchronous dynamic
decoder has a novel design where it has two fronts, with one
performing ``exploration'' and the other ``backfill''. The computation of the two fronts alternates in the decoding process, resulting in more effective pruning than the standard one-pass decoding with an on-the-fly composition decoder. Experiments show that the proposed decoder works notably faster than the standard one-pass decoding with on-the-fly composition decoder, while the acceleration will be more obvious with the increment of data complexity.
\end{abstract}


\section{Introduction}
\label{sec:intro}

Automatic speech recognition (ASR) technologies have been widely and successfully applied in many real-world fields with recent advances in deep learning algorithms, thanks to the availability of ever increasing computational power.
In particular, acoustic model (AM) inference and decoding are the main computing consumption parts of an ASR system. Researchers have proposed a variety of efficient methods to speedup acoustic model inference, including novel acoustic structures~\cite{peddinti2017low}, frame-skipping~\cite{pundak2016lower} and  quantization~\cite{mcgraw2016personalized,xiang2017binary}.

At the same time, the decoding technology is also constantly developing. In the decoding field, the core problem is how to generate an accurate lattice. Having a high quality lattice allows post-processing steps to further improve performance, e.g. lattice rescoring. A common assumption underlying lattice generation methods is the word-pair assumption. In~\cite{ortmanns1997word}, a tree-based word graph generation method is proposed to generate the lattice. In the recent decade, the weighted finite-state transducer (WFST) based lattice generation method is applied to decoders~\cite{mohri2002weighted}. In~\cite{ljolje1999efficient}, the decoder is expanded down to the context-dependent phone level (i.e. CLG). After that, the algorithm in~\cite{povey2012generating} is applied to WFSTs, expanded it down to the context-dependent state level (i.e. HCLG), which store the information of scores and state-level alignments.
Normally, the larger the language model (LM) is employed, the more accurate the lattice will be generated. Nevertheless, because of the limitation of memory, we commonly perform a two-pass decoding, where in the first pass we generate lattices with a small (e.g. low-order or pruned) LM, and the lattice is re-scored with a relatively big LM in the second pass~\cite{stolcke1997explicit,xu2018pruned}. But the two-pass procedure makes the latency issue unavoidable.
To overcome it, a useful approach is to perform one-pass on-the-fly (also called on-demand) composition decoding, in which a decoding graph with a small LM is created and then composed with a graph representing the difference between a large LM and the small LM as needed, in order to generate the search space dynamically during decoding. However, the decoding speed decreases as the search space is not as optimized as offline decoding. Researchers present many algorithms to speed it up, such as pruning~\cite{hori2004fast,nolden2012search}, look-ahead~\cite{soltau2002efficient,soltau2009dynamic,baek2001memory} and on-the-fly hypothesis re-scoring idea under \textit{phone-pair assumption}~\cite{hori2003improved}. Unfortunately, it is still difficult to generate an exact lattice with a huge LM in a fast way, so it is worthy to explore faster approaches.

The paper proposes a novel method to optimize the on-the-fly composition decoding for exact lattice generation~\cite{povey2012generating}. The proposed WFST-based decoder is denoted as \textit{asynchronous dynamic decoder} (AsyncBigLM decoder). The core novel design of the proposed decoder is that it has two fronts, with one performing ``exploration'' and the other ``backfill''. An A* algorithm is employed to evaluate which tokens worth to be back-filled.
By evaluating the proposed asynchronous dynamic decoder, we observe up to 20.17\% relative speedup.
This work is open-sourced under Kaldi~\cite{povey2011kaldi}. It is a general-purpose decoder, which does not have any special requirements for AMs or LMs, and will be compatible with all released Kaldi recipes.

\vspace{-3mm}
\section{WFST-Based Decoder}
\subsection{Basic Decoder}
\label{sec:decoder}
In Kaldi, the standard decoding graph being used is very close to~\cite{mohri2002weighted}, where the WFST decoding graph is
\begin{equation}
  \setlength\abovedisplayskip{5pt}
  \setlength\belowdisplayskip{5pt}
  S \equiv HCLG = \min(\det(H \circ C \circ L \circ G)),
\end{equation}
where \textit{H}, \textit{C}, \textit{L}, \textit{G} represent the Hidden Markov Model (HMM) structure, phonetic context-dependency, lexicon and grammar respectively, and $\circ$ represents the \textit{composition} operation of WFSTs. On an arc in \textit{HCLG}, the input label is the identifier of a clustered context-dependent HMM state, the output label corresponds to a word, and the weight typically represents a negated log-probability. A special symbol $\epsilon$ may occur on both input and output labels, which means ``no label is present''.

When we want to decode an utterance of $T$ frames, we use the \textit{token passing} algorithm~\cite{young1989token} on the HCLG graph. We denote the acoustic log-likelihood graph as $U$, which is generated by the acoustic model. The algorithm can be regarded as composing $U$ with the \textit{HCLG} graph. We denote the result of the composition as $W$, which is the searching graph of the utterance.
\begin{equation}
  \setlength\abovedisplayskip{5pt}
  \setlength\belowdisplayskip{5pt}
  W \equiv U \circ HCLG.
\end{equation}

If an exact search is performed, then $W$ would have approximately $T+1$ times more states than HCLG. After the composition, we find the best path (i.e. the path which has the lowest cost) in the searching graph. In practice, because of the time and memory limitations, the beam pruning~\cite{van1996adaptive} is performed instead of an exact search:
\begin{equation}
  \setlength\abovedisplayskip{5pt}
  \setlength\belowdisplayskip{5pt}
  P = \text{prune}(W, \alpha),
\end{equation}
where \textit{P} is the pruned graph and $\alpha$ is the beam value. During the pruning operation, we discard all paths that are not within the beam $\alpha$ compared to the best cost.

In fact, a token, which is indexed by \textit{HCLG-state} at each frame step, represents the potential decoding information of an input utterance up to the current frame. We record it as (\textit{frame-index}, \textit{HCLG-state}) pair. For each token, we keep the information of acoustic cost, graph cost and extra cost which indicate the difference between the best path through current token and the absolute best path under the assumption that any currently active states at decoding front can eventually succeed. Acoustic cost and graph cost are stored separately so that re-scaling and rescoring with higher-order LM subsequently are convenient.

\subsection{BigLM Decoder}
\label{sec:biglmdecoder}
The basic idea of the on-the-fly composition decoder ~\cite{hori2004fast}, denoted as \textit{BigLM decoder}, is to create the decoding graph HCLG with a small LM, and compose it with a WFST representing the difference between a large LM and the small LM dynamically. Imagine that the small LM is $G$, and the large one is $G'$. The decoding graph can be regarded as a two-stage composition:
\begin{align}
\setlength\abovedisplayskip{5pt}
\setlength\belowdisplayskip{5pt}
    F = -G \circ G',
    \\
    S_{big} = HCLG \circ F,
\end{align}
where $-G$ has the same topology as $G$ but with its weights negated. We refer to $F$ as \emph{residual grammar}. In practice, we keep the \textit{HCLG} and $F$ separately. Compared with the basic decoder, we keep the 3-tuple (\textit{frame-index}, \textit{HCLG-state}, \textit{F-state}) for each token. At each time step, the token passing executes the following steps:
\begin{enumerate}
    \item Get the \textit{HCLG-state} of a token, pass it on for one step in the \textit{HCLG} graph, record the output label on that arc and obtain a new \textit{HCLG-state'}.
    \item Get the \textit{LM-state} of the token, regard the output label as input label and pass it on for one step in the residual grammar (i.e.  \textit{F}). Obtain a new \textit{LM-state'}.
    \item Generate the new 3-tuple with the new \textit{HCLG-state'} and \textit{LM-state'}. The frame index depends on the input label (i.e.$\epsilon$ or not) in the \textit{HCLG} graph.
\end{enumerate}

The BigLM decoder is memory-efficient so that richer LM knowledge can be involved in one-pass decoding. Compared with the lattice re-scoring method with the same beam, the BigLM decoder usually gives better accuracy. The benefit comes from better pruning, where the information of the large LM is included, and the Viterbi beam pruning is done with closer-to-optimal language model probabilities. But the BigLM decoder is slow due to the computational overhead introduced by composition during decoding.

\section{Improved Asynchronous BigLM Decoder}
\label{sec:Astarbiglmdecoder}
\subsection{Motivation}
\begin{figure}[!ht]
  \centering
  \includegraphics[width=0.48\textwidth]{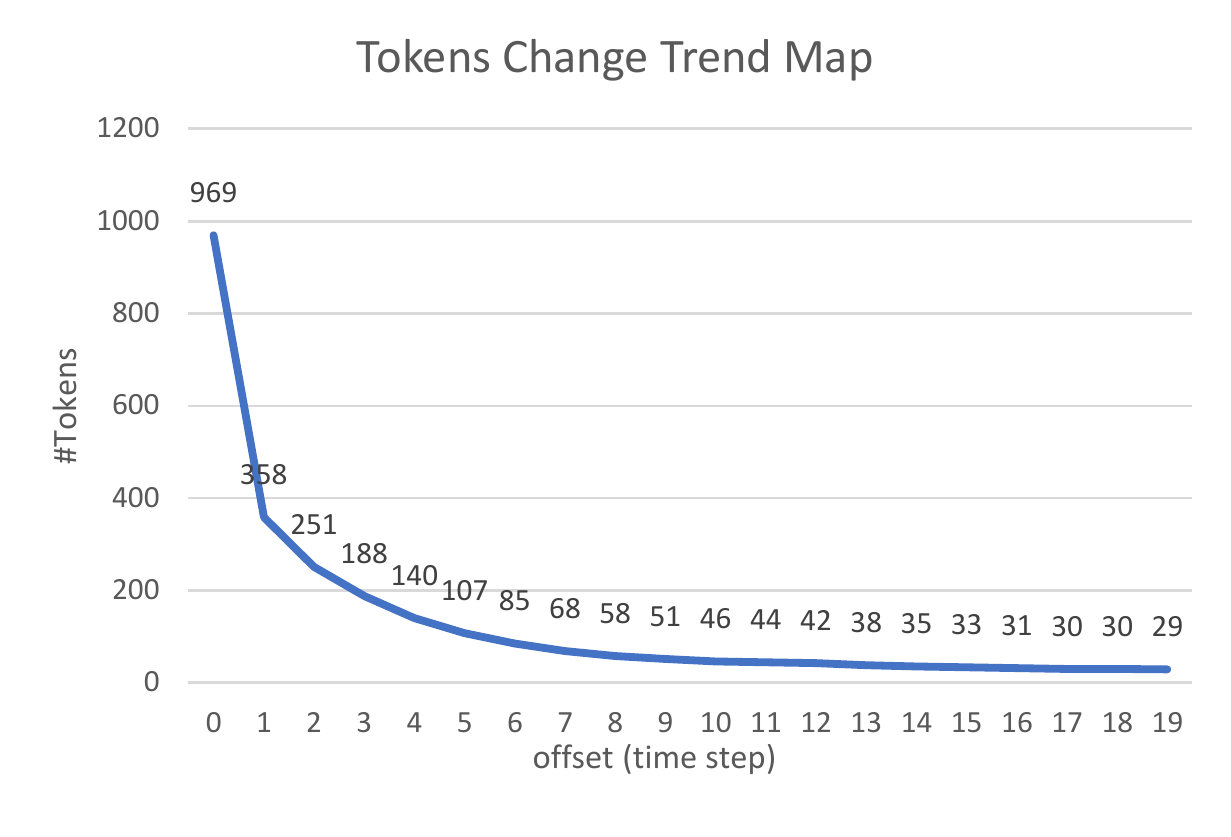}
  \vspace{-25pt}
  \caption{The trend map of tokens change. It is an illustration in the case of a WFST-based decoder. The number of tokens decreases exponentially and levels off gradually.}
  \label{fig:1}
\end{figure}
The big-picture of the proposed work is as follows: periodically, the decoder will treat tokens on the current frame as final, and prunes the state-level lattice to the lattice-beam -- discarding tokens whose costs are worse than the best one by a margin. A lot of tokens will be discarded so that the previous generation operations associated with these discarded tokens are wasted during decoding. These operations can be saved. This can be illustrated in Figure~\ref{fig:1}, where we report the number of tokens still alive at frame $(\textrm{t}-\textrm{offset})$ \footnote{$\textrm{offset} = 0 \ldots T_{1}$}, on a lattice generated from a randomly picked example from the Librispeech dataset~\cite{panayotov2015librispeech}. As the figure indicates, the number of active tokens reduces rapidly when future information is used during pruning.

\subsection{AsyncBigLM decoder}
Firstly, let's re-visit the A* method briefly.
\begin{equation}
\setlength\abovedisplayskip{5pt}
\setlength\belowdisplayskip{5pt}
    H^{*}(s) = f(s) + g^{*}(s),
\label{eq1}
\end{equation}
where $H^{*}(s)$ is the estimated score of the best complete path through state $s$. $f(s)$ is the score from the beginning to the state $s$ in the partial path, which can be obtained by accumulating the acoustic and LM probabilities during decoding straight-forwardly. $g^{*}(s)$ is an estimate of the best partial path from state $s$ to the end. The key to the A* method is how to estimate a reasonable $g^{*}(s)$.

Our proposed AsyncBigLM decoder is similar to the BigLM decoder in which its searching space is constructed in (\textit{frame-index}, \textit{HCLG-state}, \textit{F-state}) space. Nevertheless, it works in a different way where it has two ``decoding fronts'', namely ``exploration'' front and ``backfill'' front respectively. The former one occurs at the current frame $t$ and the latter one at frame $t-\textrm{offset}$~\footnote{``offset'' is set up empirically.}. Basically, we process the ``best-in-class'' token, which has the best cost for each specific HCLG-state $s$ on the ``exploration'' front. The ``not-best-in-class'' tokens will be processed on the ``backfill'' front. We deal with the two fronts alternately, so the sequence will be something like: explore for frame $t$, backfill for frame $(t-\textrm{offset})$, explore for frame $(t+1)$, backfill for frame $(t+1-\textrm{offset})$, $\ldots$ and so on~\footnote{The process on each front can be executed for a few frames to balance the accuracy and speed. E.g. explore for frame $t, t+1$; backfill for frame $(t-\textrm{offset}), (t+1-\textrm{offset})$; explore for frame $t+2, t+3$; backfill for frame $(t+2-\textrm{offset}), (t+3-\textrm{offset})$; \ldots and so on.}. The details of these two fronts are described in the following two sections.

\begin{figure}[!ht]
  \centering
  \includegraphics[width=0.45\textwidth]{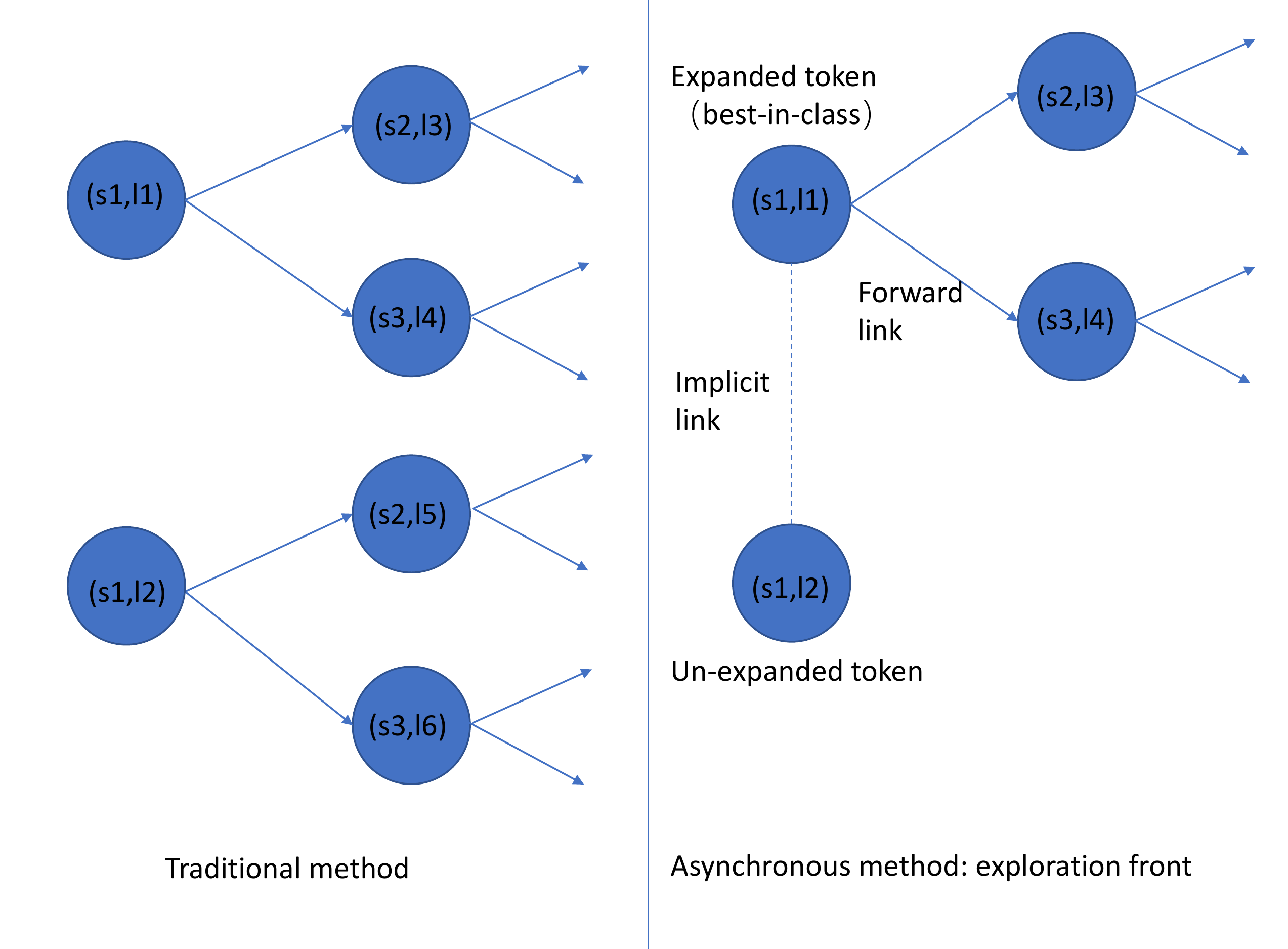}
  \caption{The difference between traditional biglm decoding and improved biglm decoding. In the traditional case, each state pair is expanded in current frame. But in the improved case, only the best token for each HCLG state is expanded.}
  \label{fig:2}
\end{figure}

\vspace{-6mm}
\subsubsection{Exploration Front}
We show the difference between BigLM decoder and the AsyncBigLM decoder in the forward pass in Fig~\ref{fig:2}. Let's write the state-pair (\textit{HCLG-state}, \textit{F-state}) on each frame as (s, l) for short. On the exploration front at frame $t$, the operations would be similar to the current BigLM decoder, with one exception -- suppose there are some tokens having the same \textit{HCLG-state} and different \textit{F-state}s on the current frame (e.g. the (s1, l1) and (s1, l2) tokens in Fig.~\ref{fig:2}). In the BigLM decoder, all arcs leaving each state should be processed, but in the AsyncBigLM decoder, only the ``best-in-class'' token, which has the best forward cost (the $\alpha$ cost in the HMM sense) in all tokens with the same \textit{HCLG-state} on the frame, will be expanded on the exploration front (e.g. Token (s1, l1) in Fig.~\ref{fig:2}). It is also called as ``expanded'' token. Then the ``not-best-in-class'' tokens, denoted as ``un-expanded tokens'', will not be expanded immediately (e.g. Token (s1, l2) in Fig.~\ref{fig:2}). Instead we create \textit{implicit links} from them to the expanded token. We will consider processing all the un-expanded tokens on the backfill front. In conclusion, the procedure above would suppress the propagation of all but the ``best-in-class'' token for each \textit{HCLG-state}.

\subsubsection{Backfill Front}
\begin{figure}[!htp]
  \centering
  \includegraphics[width=0.35\textwidth]{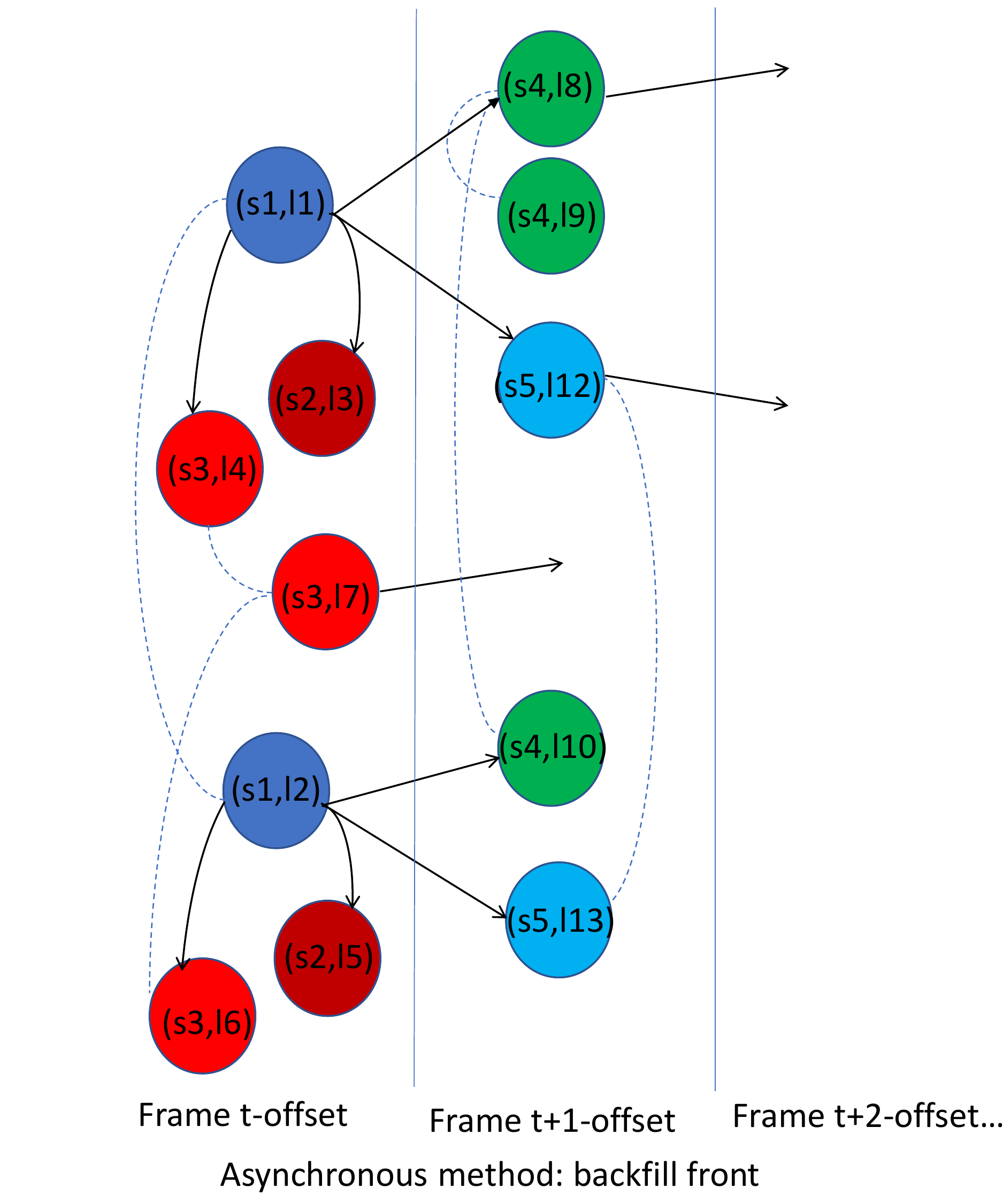}
  \vspace{-15pt}
  \caption{Backfill front: The token with same colour has the same \textit{HCLG-state} but different \textit{F-states}. The ``un-expanded'' token follows the ``expanded'' token to expand itself. (The blue dotted lines represent the \textit{implicit links} and the black arrow lines represent the \textit{forward links}.)}
  \vspace{-25pt}
  \label{fig:3}
\end{figure}

On the backfill front, we expand the previous un-expanded tokens. Firstly, we need to decide which un-expanded tokens should be processed with the help of the A* method. We treat the expanded tokens on the exploration front as final and compute the backward cost (the $\beta$ cost in the HMM sense) of all the expanded tokens from the exploration front to the backfill front. Then we assume that the $g^{*}(s)$ of the un-expanded tokens are the same as the $g^{*}(s)$ of the expanded tokens that these un-expanded tokens link to (e.g. Token (s1, l2) can borrow the $g^{*}(s)$ of Token (s1, l1) in Fig.~\ref{fig:2}). So we can obtain $H^{*}(s)$ of each un-expanded token with from $f(s)$ and the $g^{*}(s)$ as shown in Eq.(~\ref{eq1}). Then $H^{*}(s)$ of each un-expanded token is compared with that of the best token to decide whether the token should be expanded on the backfill front.

When we expand an previous un-expanded token, we follow the footsteps of the arcs of the corresponding expanded token, using only the information present in the \textit{forward links} leaving the expanded token, i.e., not revisiting the graph and the acoustic likelihoods. When we create new tokens on the destination-states, the implicit links of them are set. These implicit links will be used to expand these destination tokens appropriately when we do back-fill in the next time. The situation is illustrated in Fig~\ref{fig:3}. For example, the un-expanded token (s1,l2) is expanded by following the expanded token (s1,l1)'s footsteps and the implicit links of all destination tokens are set.

In addition, when processing these previously un-expanded tokens, we need to pass the information from the backfill front to the exploration front immediately in two special cases:
\\
1) On the backfill front, when we expand a token and its destination token reaches an ``existing state'' (one that was created during a prior exploration step, for instance), and the destination token has a better forward cost than the token on existing state. In this case, before further exploration, we would propagate the cost change along all the paths starting from the existing state, so that in the future exploration we can decode with up-to-date forward-costs. Otherwise the exploration forward-cost would be permanently ``out-of-step''.
\\
2) A new token is created on state (s, l), and it has better cost than the existing tokens on the same \textit{HCLG-state} $s$. In this case, before the further ``exploration front'' is performed, we expand the new token till the current frame -- this guarantees we process the correct ``best-in-class'' tokens on the exploration front.

\vspace{-6mm}
\section{Experiment}
\label{sec:exp}
\vspace{-2mm}
We use the open-source speech recognition toolkit Kaldi~\cite{povey2011kaldi} to conduct the experiments. We evaluate the algorithm with the corpus--LibriSpeech~\cite{panayotov2015librispeech} (LIB) which contains about 960 hours training data and 4 kinds of separate test data-sets (dev-clean,  dev-other, test-clean and test-other). Each of them has about 2 hours audio data.

All the acoustic models are trained with TDNN~\cite{peddinti2015time} structure and lattice-free maximum mutual information~\cite{povey2016purely} (LF-MMI) criterion. Besides our improved AsyncBigLM decoder, we employ two methods as baselines. The first one is ``lattice-rescoring'' method which generates the lattices with a small LM and rescoring them with a large LM. The second one is the BigLM decoder which is described in Section~\ref{sec:biglmdecoder}. For the LibriSpeech testing, the customary standard LMs are used~\cite{panayotov2015librispeech}. The small 3-gram LM (60MB) is employed to build the HCLG graph. The mild-pruned 3-gram LM (tgmed, 140MB), original 3-gram LM (tglarge, 760MB) and original 4-gram LM (fglarge) are used to build \textit{residual grammar} separately.

\begin{figure}[!htp]
  \centering
  \vspace{-10pt}
  \includegraphics[width=0.48\textwidth]{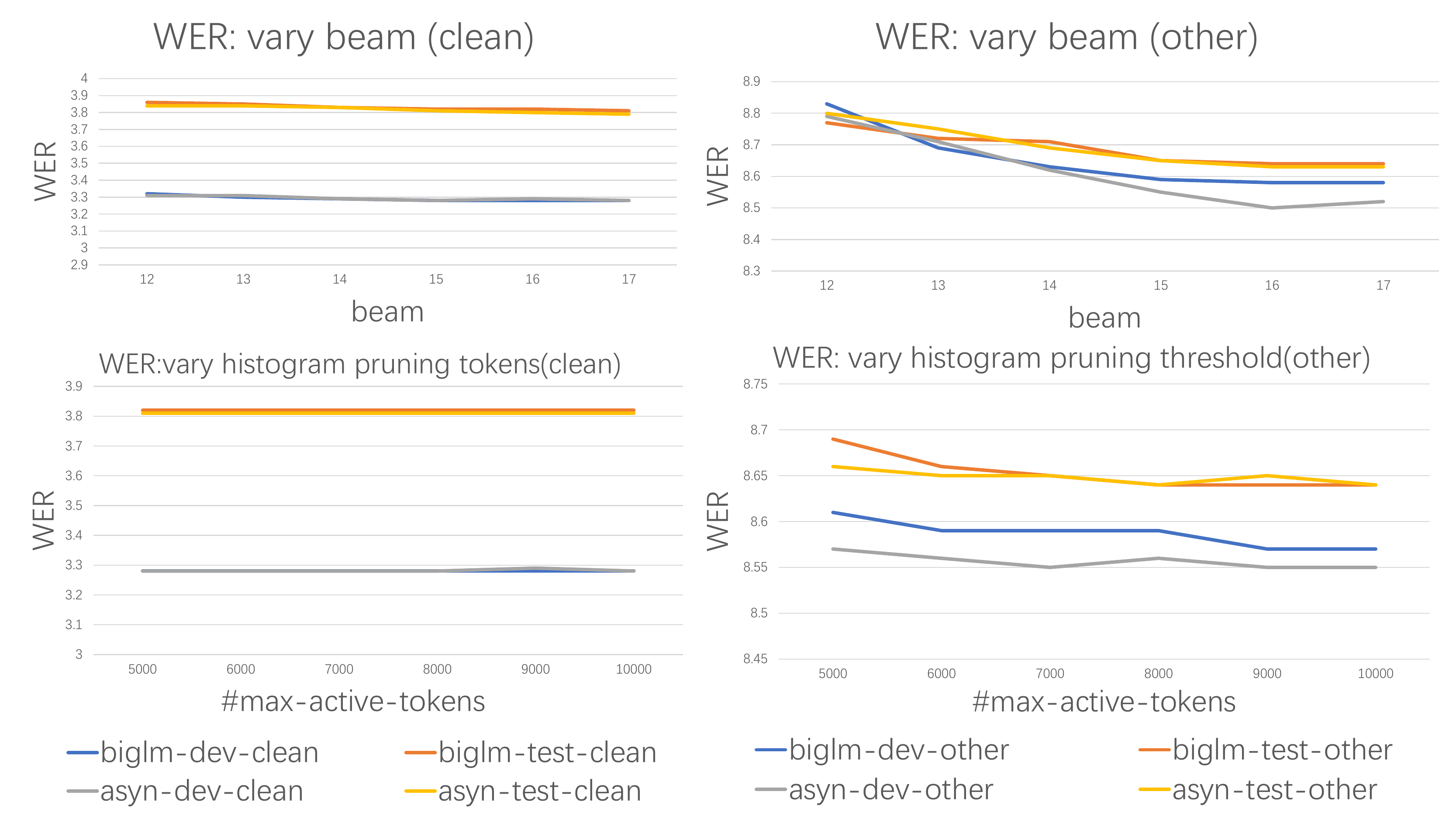}
  \vspace{-15pt}
  \caption{The WER between BigLM and AsyncBigLM. It shows the performance of the
  two kinds of decoders are similar}
  \label{fig:4}
  \vspace{-15pt}
\end{figure}

\begin{figure}[t]
  \centering
  \includegraphics[width=0.48\textwidth]{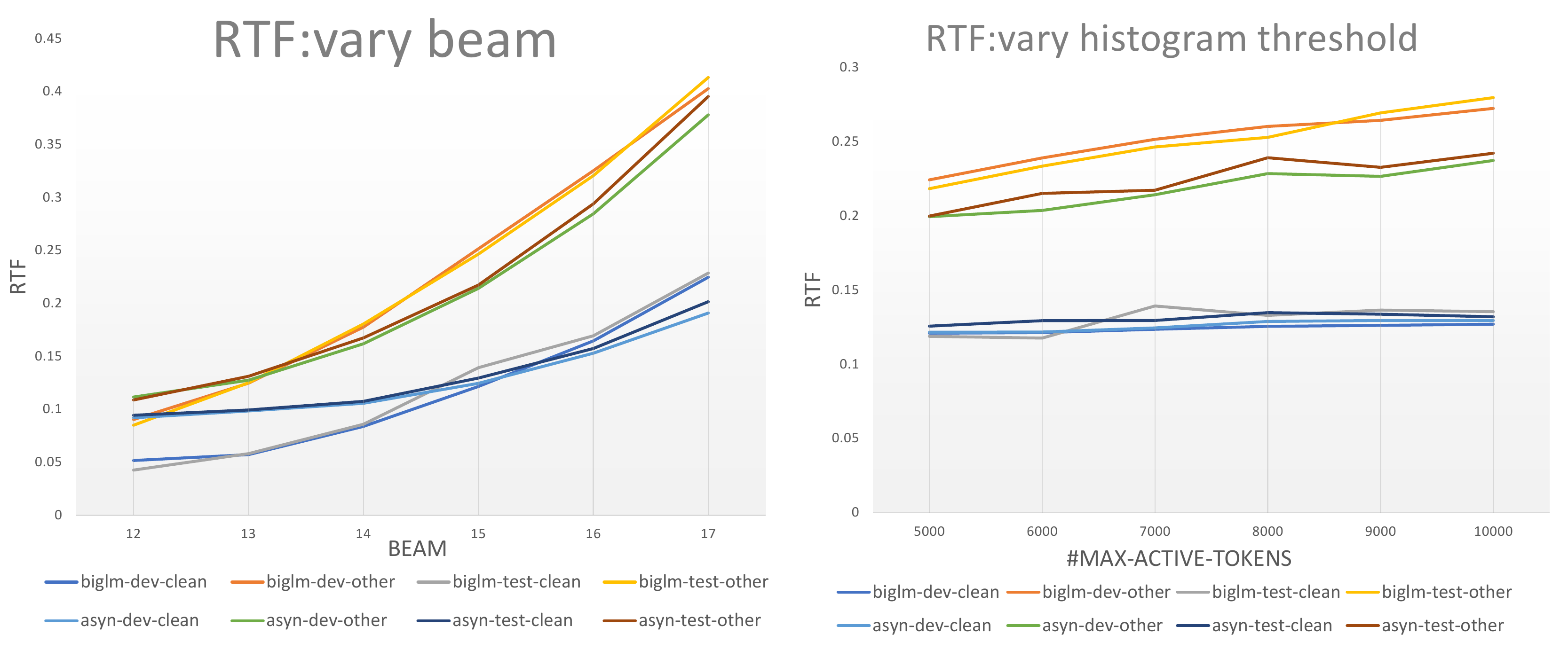}
  \vspace{-15pt}
  \caption{The RTF between BigLM and AsyncBigLM. The latter one is better than
  the former when the searching graph is large}
  \label{fig:5}
  \vspace{-20pt}
\end{figure}

In the following, unless otherwise specified, we show the results with the histogram pruning threshold (maximum-active-tokens=7000), beam pruning (beam=15) and lattice pruning (lattice-beam=8) for space reason. As shown in Fig~\ref{fig:4} and Fig~\ref{fig:5}, we tried to tune the hyper-parameters and the trend of results is similar.

\vspace{-3.5mm}
\subsection{Accuracy}
\begin{table}[!htp]
\footnotesize
    \centering
    \vspace{-23pt}
    \caption{WER statistics: Rescoring/BigLM/AsyncBigLM}
    \begin{tabular}{|c|ccc|}
    \hline
    & tgmed & tglarge & fglarge \\
    \hline
    dev-clean & 4.27/4.25/4.24 & 3.38/3.38/3.38 & 3.27/3.28/3.28 \\
    dev-other & 11/11/11.04 &
    9.14/9.1/9.1 & 8.7/8.59/8.55 \\
    test-clean & 4.74/4.77/4.77 &
    3.94/3.93/3.92 & 3.83/3.82/3.81 \\
    test-other & 11.2/11.18/11.18 &
    9.21/9.17/9.15 & 8.72/8.65/8.65 \\
    \hline
    \end{tabular}
    \label{tab1}
    \vspace{-10pt}
\end{table}

In speech recognition task, the most straight-forward evaluation criterion is word error rate (WER). In Table 1, we compare the WER among ``Lattice-rescoring'', BigLM decoder and our proposed AsyncBigLM decoder. It shows that WERs of the three methods are close, but the last two methods are better than the first a little bit. The benefit comes from the fact that the information of the large LM is included when the decoder goes across the word boundary. Compared with WERs, average log-likelihoods of lattices will provide more accurate information for decoder evaluation. Thus in Table 2, we compare the average log-likelihood between BigLM decoder and AsyncBigLM decoder. As ``lattice-rescoring'' is generated from the small graph, the average log-likelihood is far behind. From Table 2, we can see that the differences are extremely small ($<0.0001$), which means the accuracy of the two decoders is similar under the same condition. As the ``BigLM''-kind decoders are better, we compare them further. In Figure~\ref{fig:4}, we show the results when we tune the decoding beam/histogram pruning threshold. We can see that the WERs are comparable.

\begin{table}[!htp]
\footnotesize
    \centering
    \vspace{-18pt}
    \caption{Log-likelihood statistics: BigLM/AsyncBigLM}
    \begin{tabular}{|c|ccc|}
    \hline
    & tgmed & tglarge & fglarge \\
    \hline
    dev-clean & 3.80056/3.80054 &
    3.83091/3.83085 & 3.84251/3.84245 \\
    dev-other & 3.32441/3.32439 &
    3.34851/3.34839 & 3.3593/3.35919 \\
    test-clean & 3.73661/3.73661 &
    3.76555/3.76551 & 3.77599/3.77594 \\
    test-other & 3.2984/3.2984 &
    3.32344/3.32342 & 3.33388/3.33387 \\
    \hline
    \end{tabular}
    \label{tab2}
    \vspace{-13pt}
\end{table}

\vspace{-3.3mm}
\subsection{Speedup}

\begin{table}[!htp]
\footnotesize
    \centering
    \vspace{-27pt}
    \caption{RTF: Biglm/AsyncBiglm}
    \begin{tabular}{|c|ccc|}
    \hline
    & tgmed & tglarge & fglarge \\
    \hline
    dev-clean & 0.1178 / 0.1056 & 0.1158 / 0.1141 & 0.1216 / 0.1233 \\
    dev-other & 0.1998 / 0.1877 & 0.2165 / 0.2061 & \textbf{0.2517 / 0.2094} \\
    test-clean & 0.1019 / 0.1106 &  0.1242 / 0.1188 & \textbf{0.1394 / 0.1294} \\
    test-other & 0.2075 / 0.1937 & \textbf{0.2366 / 0.2017} & \textbf{0.2466 / 0.2172} \\
    \hline
    \end{tabular}
    \vspace{-10pt}
    \label{tab3}
\end{table}

Under the same pruning parameters condition, we compare the real time factor (RTF) between BigLM and AsyncBigLM to show the speedup performance in Table~\ref{tab3}. The bold figures in Table 3 show the speedup rate is about $7.6-20.17\%$. While as shown in Figure~\ref{fig:5}, the RTFs of AsyncBigLM are better than BigLM gradually as we increase the decoding beam or histogram pruning threshold. Table~\ref{tab3} also reflects the trend that as the increase of data complexity (i.e. noisy data or bigger LM), the speedup improvement is more obvious. We believe the reasons are that: 1) For noisy data, it might lead to less discriminative likelihoods from the acoustic model so that more hypotheses with similar scores need to be processed in BigLM decoder. 2) For the bigger LM, it will lead to a bigger residual grammar space. However, as the AsyncBigLM only propagates the best-in-class tokens on the exploration front and skip unpromising tokens on the backfill front, it apparently saves lots of operations from the two aspects so that the speed increase more obvious.

As the expectation in Section 4.1, the AsyncBigLM decoder achieves the benefit from saving searching operations. We count the effective propagation times on both BigLM decoder and AsyncBigLM decoder. The propagation statistics comparison further shows the reasons of acceleration: 1) the propagation times of asynchronous decoder is significantly less than that of the BigLM decoder on the exploration front; 2) the propagation times on the backfill front is limited and the total times of AsyncBigLM decoder is still less than the BigLM decoder's. They prove the validity of our proposed asynchronous method.

\begin{table}[!htp]
\footnotesize
    \centering
    \vspace{-20pt}
    \caption{The effective propagation times (million)}
    \begin{tabular}{|c|cc|}
    \hline
    & Exploration & Backfill \\
    \hline
    BigLM & 5.97 & 0 \\
    AsyncBigLM & 3.84 & 0.3 \\
    \hline
    \end{tabular}
    \label{tab4}
    \vspace{-18pt}
\end{table}

\vspace{-3mm}
\section{Conclusion}
\label{sec:conclusion}
\vspace{-3mm}
In this paper, we proposed a smart AsyncBigLM decoder with A* method to speedup the one-pass on-the-fly composition decoding. The proposed algorithm can achieve up to $20.17\%$ speedup rate. More importantly, the speedup would be more prominent and stable as the complexity of data increases.

\bibliographystyle{IEEEbib}
\bibliography{refs}

\end{document}